\documentclass[aps,prb,twocolumn,superscriptaddress,showpacs]{revtex4-1}
\usepackage{graphics}
\usepackage{subfigure}
\usepackage{longtable}
\usepackage{graphicx}
\usepackage{pstricks}
\usepackage{amsmath}
\usepackage{dcolumn}
\usepackage{bm}
\usepackage{pdfpages}

\begin{document}

\title{Unified picture of the doping dependence of superconducting transition temperatures in alkali metal/ammonia intercalated FeSe}

\author{Daniel Guterding}\email{guterding@itp.uni-frankfurt.de}\author{Harald O. Jeschke}
\affiliation{Institut f\"ur Theoretische Physik, Goethe-Universit\"at Frankfurt, 
Max-von-Laue-Stra{\ss}e 1, 60438 Frankfurt am Main, Germany}

\author{P. J. Hirschfeld}
\affiliation{Department of Physics, University of Florida, Gainesville, Florida 32611, U.S.A.}

\author{Roser Valent\'\i }
\affiliation{Institut f\"ur Theoretische Physik, Goethe-Universit\"at Frankfurt, 
Max-von-Laue-Stra{\ss}e 1, 60438 Frankfurt am Main, Germany}

\begin{abstract}
In the recently synthesized Li$_x$(NH$_2$)$_y$(NH$_3$)$_z$Fe$_2$Se$_2$ family 
of iron chalcogenides a molecular spacer consisting of lithium ions, lithium 
amide and ammonia separates layers of FeSe. It has been shown that upon 
variation of the chemical composition of the spacer layer, superconducting 
transition temperatures can reach $T_c\sim 44~\mathrm{K}$, but the relative 
importance of the layer separation and effective doping to the $T_c$ 
enhancement is currently unclear. Using state of the art band structure 
unfolding techniques, we construct eight-orbital models from \textit{ab-initio} 
density functional theory calculations for these materials. Within an RPA 
spin-fluctuation approach, we show that the electron doping enhances the 
superconducting pairing, which is of $s_\pm$-symmetry and explain the 
experimentally observed limit to $T_c$ in the molecular spacer intercalated 
FeSe class of materials.
\end{abstract}

\pacs{
71.15.Mb, 
71.18.+y, 
71.20.-b, 
74.20.Pq, 
74.24.Ha, 
74.70.Xa   
}

\maketitle

After the discovery of iron based superconductors in 2008,
transition temperatures were quickly improved to $\sim 56~\mathrm{K}$
by chemical substitution~\cite{SrFeAsF56K}. Recently, the possible
discovery of superconductivity with $T_c= 65~\mathrm{K}$~\cite{Tang2013}
and even $T_c \sim 100~\mathrm{K}$~\cite{FeSeSTO100K}  in single-layer FeSe films grown
by molecular beam epitaxy on SrTiO$_3$ showed that temperatures close to and above
the boiling point of liquid nitrogen ($77~\mathrm{K}$)  might be achievable.
These results have initiated an intensive debate regarding the origin of
the high superconducting temperatures and the role played by
electron doping via substrate, dimensionality and lattice strain.

While bulk FeSe has a $T_c$ of only 8-$10~\mathrm{K}$,  it has been known for some time that 
it can be substantially enhanced, to $40~\mathrm{K}$ or higher by alkali intercalation~\cite{KFe2Se2discovery}.
Materials with a single alkali A = K, Cs, Rb between FeSe layers of nominal form  A$_x$Fe$_{2-y}$Se$_2$ have been intensively studied,
and shown to display a wide variety of unusual behaviors relative to the Fe pnictide superconducting 
materials~\cite{DagottoReview2013}. These include likely phase separation into an insulating 
phase with block antiferromagnetism and ordered Fe vacancies, and a superconducting phase that is 
strongly alkali deficient and whose Fermi surface as measured by ARPES apparently contains no 
holelike Fermi surface pockets, in contrast to Fe-pnictides. Since the popular spin fluctuation scenario for $s_\pm$
pairing relies on near nesting of hole and electron pockets, it has been speculated that a different mechanism
for pairing might be present in these materials, and even within the spin fluctuation approach, different gap 
symmetries including $d$-wave pairing have been proposed~\cite{KFeSeFRG,KFeSeMaier,GapEvolutionMaiti,KFeSeKreisel}.  
The gap symmetry and structure is still controversial, however~\cite{Xu2012,Mazin2011}.
      
\begin{figure}[tbp]
\centering

\includegraphics[width=0.35\linewidth]{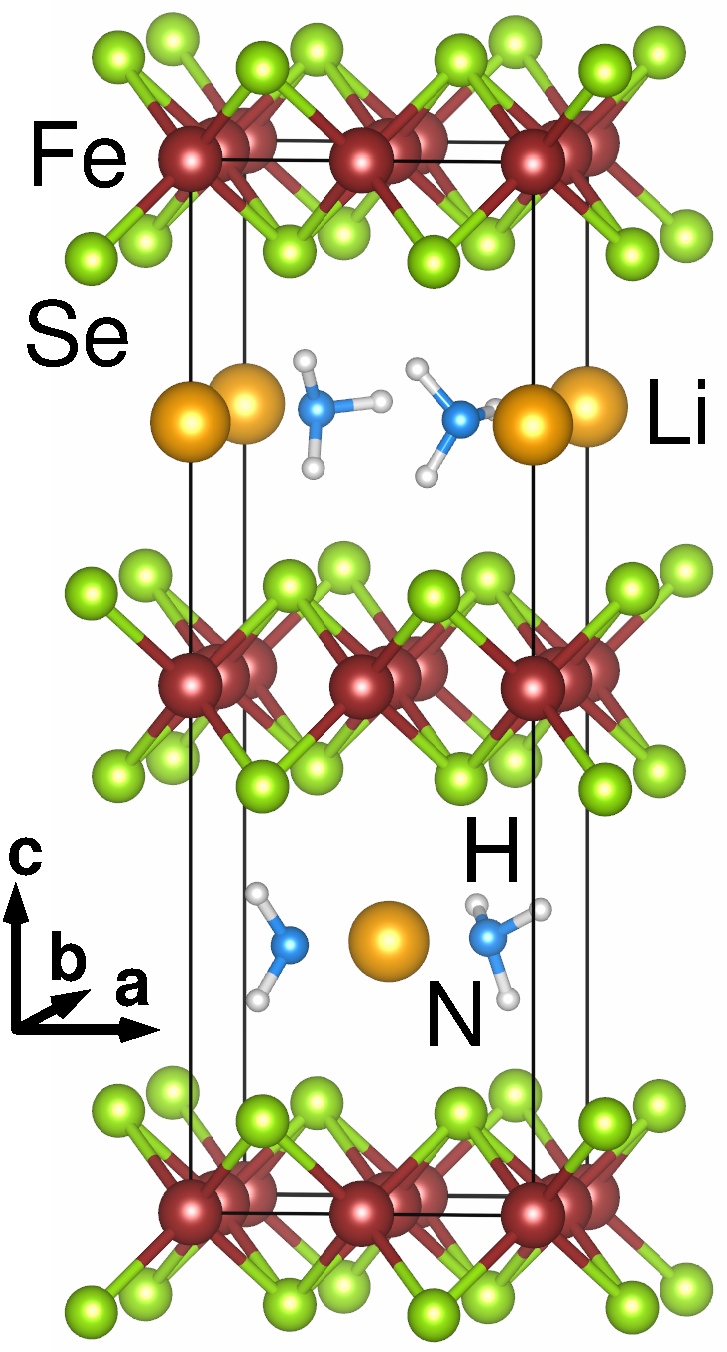}

\caption{(Color online) Idealized crystal structure of Li$_{0.5}$(NH$_3$)Fe$_2$Se$_2$. 
For a detailed discussion of experimental crystal structures
see Ref. \onlinecite{AmmoniaRichJACS, AmmoniaPoorNature}.}
\label{fig:crystalstructure}
\end{figure}

In addition to the unusual doping, speculation on the origin of the higher $T_c$ has centered on
the intriguing possibility that enhancing the FeSe layer spacing improves the two-dimensionality of the band
structure and hence Fermi surface nesting~\cite{ScheidtPureAmmonia,IntercalateLayerDistance}.  
In an effort to investigate this latter effect, organic molecular complexes including alkalis were recently 
intercalated between the FeSe layers~\cite{AmmoniaPoorNature, AmmoniaRichJACS,
LithiumIronHydroxideSelenides, IntercalateLayerDistance,
AlkalineEarthIntercalates, ScheidtPureAmmonia,
PyridineIntercalate, SodiumIntercalate}, yielding transition temperatures up to 46 K. The most
intensively studied materials incorporate molecules including ammonia, 
for example Li$_{0.56}$(NH$_2$)$_{0.53}$(NH$_3$)$_{1.19}$Fe$_2$Se$_2$ with $T_c =
39~\mathrm{K}$~\cite{AmmoniaRichJACS} and
Li$_{0.6}$(NH$_2$)$_{0.2}$(NH$_3$)$_{0.8}$Fe$_2$Se$_2$ with $T_c =
44~\mathrm{K}$~\cite{AmmoniaPoorNature}. The crystal structure of a
stoichiometric version of these materials is shown in Fig. \ref{fig:crystalstructure}.  
Recently,  Noji \textit{et al.}~\cite{IntercalateLayerDistance} compared
data on a wide variety of FeSe intercalates and noted a strong
correlation of $T_c$  with interlayer spacing, corresponding to a nearly 
linear increase between 5 to $9~\mathrm{\AA}$, followed by a rough 
independence of spacing with further increase between 9 to $12~\mathrm{\AA}$.

In the present work we study the question of how exactly doping and interlayer distance influence
$T_c$ in molecular intercalates of FeSe, whether these effects are separable, and what gives rise to the
apparent upper limit for $T_c$ in this family of iron chalcogenides. Using a combination of
first principles calculations of the electronic structure of several materials in this class, together with
model calculations of spin fluctuation pairing based on these band structures, we
argue that strength and wave-vector of
spin-fluctuations in lithium/ammonia intercalated FeSe can be
controlled by tuning the Li$^+$:NH$_2^-$ ratio in the spacer layer.
We show that the evolution of $T_c$ 
with electron doping can be understood from the shape of the density
of states close to the Fermi level. As long as hole pockets are present,
we find that the superconducting
pairing is of $s_\pm$ character 
and identify a subleading $d_{x^2 - y^2}$ instability. We believe
that our interpretation is valid in a broad class of
related materials.

We performed density functional theory calculations
 for
Li$_{0.5}$(NH$_2$)$_y$(NH$_3$)$_z$Fe$_2$Se$_2$ at various ratios of
NH$_2^-$ to NH$_3$ content, starting from the experimentally
determined structures
Li$_{0.56}$(NH$_2$)$_{0.53}$(NH$_3$)$_{1.19}$~\cite{AmmoniaRichJACS}
and
Li$_{0.6}$(NH$_2$)$_{0.2}$(NH$_3$)$_{0.8}$~\cite{AmmoniaPoorNature},
which include fractionally occupied atomic sites for lithium, hydrogen
and nitrogen.
 In order to accommodate the experimental stoichiometry we
construct a $2 \times 1\times 1$ (4 Fe) supercell for the former, and
a $2 \times 2\times 1$ (8 Fe) supercell for the latter compound. We
replace all fractionally occupied nitrogen positions by fully occupied
positions. As hydrogen positions are not known precisely from
experiment, we arrange the hydrogen atoms so that we obtain NH$_3$
groups with angles of about $108^\circ$ as encountered in ammonia
and further relax these positions within the local density
approximation (LDA)~\cite{LDAPerdewWang}  with the projector augmented wave
(PAW) basis~\cite{PAWmethod} as implemented in GPAW~\cite{GPAWmethod}
 until forces are below
$2~\mathrm{meV/\AA}$. In
the $2 \times 1\times 1$ supercell we place the lithium atom in one
half of the unit cell and leave the lithium position in the other half
unoccupied. In the $2 \times 2\times 1$ supercell we arrange the
lithium atoms in a checkerboard pattern of occupied and vacant sites
(Fig. \ref{fig:crystalstructure}).

Initially, we only consider charge neutral NH$_3$ ammonia groups in the spacer
and no NH$_2^-$ .
In this way,  we obtain idealized structures with formula units
Li$_{0.5}$(NH$_3$)Fe$_2$Se$_2$ and
Li$_{0.5}$(NH$_3$)$_2$Fe$_2$Se$_2$ where chalcogen height
 and unit cell 
parameters are chosen as in the experimental
 structures~\cite{AmmoniaRichJACS,AmmoniaPoorNature}. Both structures belong to the space
group P1 because of NH$_3$ situated in the spacer layer.
Note that by setting up both structures with neutral NH$_3$,
 we are able to disentangle possible effects
of the structural differences from the effect of doping through the
composition of the spacer layer.

The experimentally
available samples~\cite{AmmoniaRichJACS,AmmoniaPoorNature}
  contain both NH$_3$ and NH$_2^-$. The radical   
 NH$_2^-$   neutralizes the
charge donated to the FeSe layer by Li$^+$ and reduces the doping
level. In order to capture  this compensation
of charge in our simulations, we use the virtual
crystal approximation (VCA)~\cite{Supplement} starting from supercells
Li$_{0.5}$(NH$_3$)Fe$_2$Se$_2$ and Li$_{0.5}$(NH$_3$)$_2$Fe$_2$Se$_2$,
which correspond to the maximally electron doped compounds.
The use of VCA has the advantage that doping is
treated in a continuous and rather isotropic, but not rigid band
fashion. We checked these calculations by removing hydrogen atoms
explicitly instead of doing VCA and found the differences to be
negligible.

\begin{figure}[tbp]
\centering

\includegraphics[width=\linewidth]{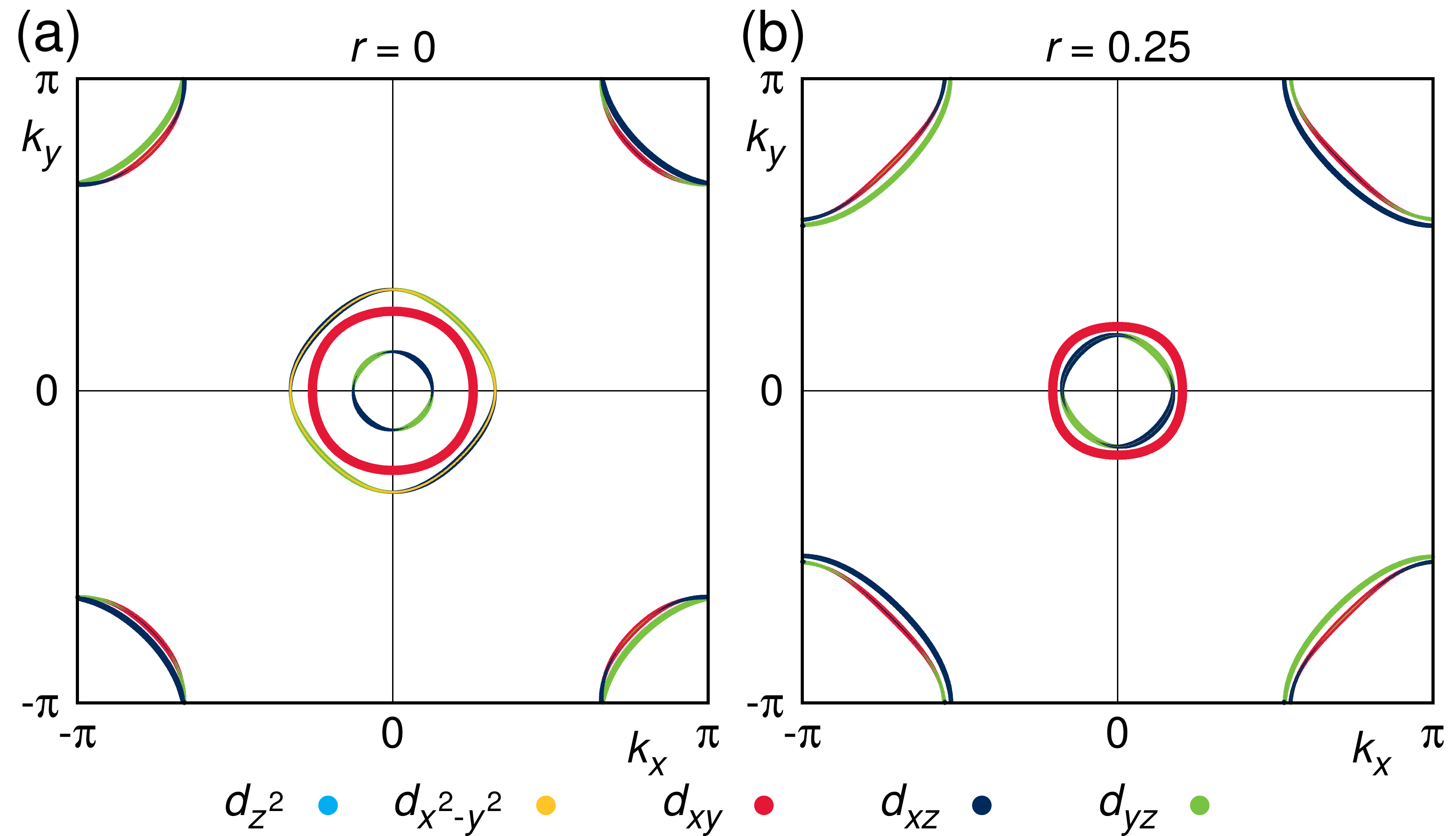}

\caption{(Color online) Fermi surface in the 16 band tight binding model for $r=0.0$
  (a) and $r=0.25$ (b) in the 2-Fe Brillouin zone at $k_z=0$. 
The colors indicate the weights of Fe 3$d$ states.}
\label{fig:16bandmodelfermisurface}
\end{figure}

The analysis of the band structure of these supercells
is done 
within an all electron full potential local orbital
(FPLO)~\cite{FPLOmethod} basis and we use LDA
 as exchange-correlation
functional~\cite{LDAPerdewWang}.  We then use projective Wannier functions as implemented in
FPLO~\cite{FPLOtightbinding} to downfold the band structure. In our
tight binding models, we keep the Fe $3d$ and Se $4p$ states. In order
to obtain band structure and Fermi surface of the supercells in the
conventional two iron unit cell, we use our recently developed
technique~\cite{Tomicunfolding} to translationally unfold the 32 and
64 band supercell models to a 16 band model of the 2 Fe equivalent
Brillouin zone. For calculations of susceptibility and superconducting
pairing, we use subsequent glide reflection
unfolding~\cite{Tomicunfolding} of the bands to obtain the 8 band
model of the 1 Fe equivalent Brillouin zone.

First we investigated the properties of the maximally electron doped
compounds in our study, Li$_{0.5}$(NH$_3$)Fe$_2$Se$_2$ (ammonia poor) and
Li$_{0.5}$(NH$_3$)$_2$Fe$_2$Se$_2$ (ammonia rich).
 Both feature at the Fermi level (not shown) two large electron
pockets in the corners of the 2 Fe Brillouin zone and two small hole
pockets around $\Gamma$. This confirms that the lithium atoms donate
electrons to the FeSe layer. Both systems have the same electron doping
but different interlayer spacing. This is observed in 
 the $k_z$-dispersion of the Fermi surface, where the
smaller interlayer distance of the ammonia poor compound leads to a
slightly increased corrugation of the cylinders.

In the experimentally realized compounds
Li$_{0.56}$(NH$_2$)$_{0.53}$(NH$_3$)$_{1.19}$
and Li$_{0.6}$(NH$_2$)$_{0.2}$(NH$_3$)$_{0.8}$
 the spacer layer nominally donates
a charge of $0.015$ and $0.2$ electrons per iron atom respectively. These doping
levels are lower than in our model materials Li$_{0.5}$(NH$_3$)Fe$_2$Se$_2$
and Li$_{0.5}$(NH$_3$)$_2$Fe$_2$Se$_2$.
To investigate the doping dependence of the electronic structure
at a given interlayer spacing,
 we consider  Li$_{0.5}$(NH$_3$)Fe$_2$Se$_2$
 and hole dope it by means of the
virtual crystal approximation as explained above.
 To simplify the notation, we label
compounds from now on not by their full chemical formula, but by an
index $r=\{ 0.0, \ldots, 0.25 \}$, which refers to the chemical formula
Li$_{0.5}$(NH$_2$)$_{0.5-2r}$(NH$_3$)$_{0.5+2r}$Fe$_2$Se$_2$.
 $r=0.25$ refers to
the compound Li$_{0.5}$(NH$_3$)Fe$_2$Se$_2$ with maximal electron
doping, where lithium nominally transfers a quarter of an electron to
each iron atom. Increasing the NH$_2^-$ content immediately brings up a third hole
pocket to the Fermi level,
 which is three-dimensional at intermediate doping and becomes
two-dimensional once the charge introduced by lithium is fully
compensated by NH$_2^-$ groups. $r=0$ denotes the compound where the charge
introduced by lithium is nominally compensated by NH$_2^-$ and no
electrons are donated to the FeSe layer.
 The Fermi surfaces
of the end members ($r=0.0$ and $r=0.25$) are shown in
Fig.~\ref{fig:16bandmodelfermisurface}. The band structure on high-symmetry paths
is included in the Supplemental Material~\cite{Supplement}.

Upon further analysis of the tight-binding parameters we find that
the hole pockets do not only shrink because the electron doping raises the
Fermi level, but also because the nearest neighbor hopping in the Fe 3$d_{xy}$
orbital decreases steeply as a function of electron doping. This near cancellation
of direct and indirect hopping paths has been discussed in the literature for 
other iron based superconductors~\cite{KotliarKineticFrustration, KurokiDiagonalHopping,AndersenBoeri}.
In the materials investigated here, we find that the degree of localization in
the Fe 3$d_{xy}$ orbital can be tuned with relatively low electron doping. Further
information are given in the Supplemental Material~\cite{Supplement}.

\begin{figure}[t]
\centering

\includegraphics[width=\linewidth]{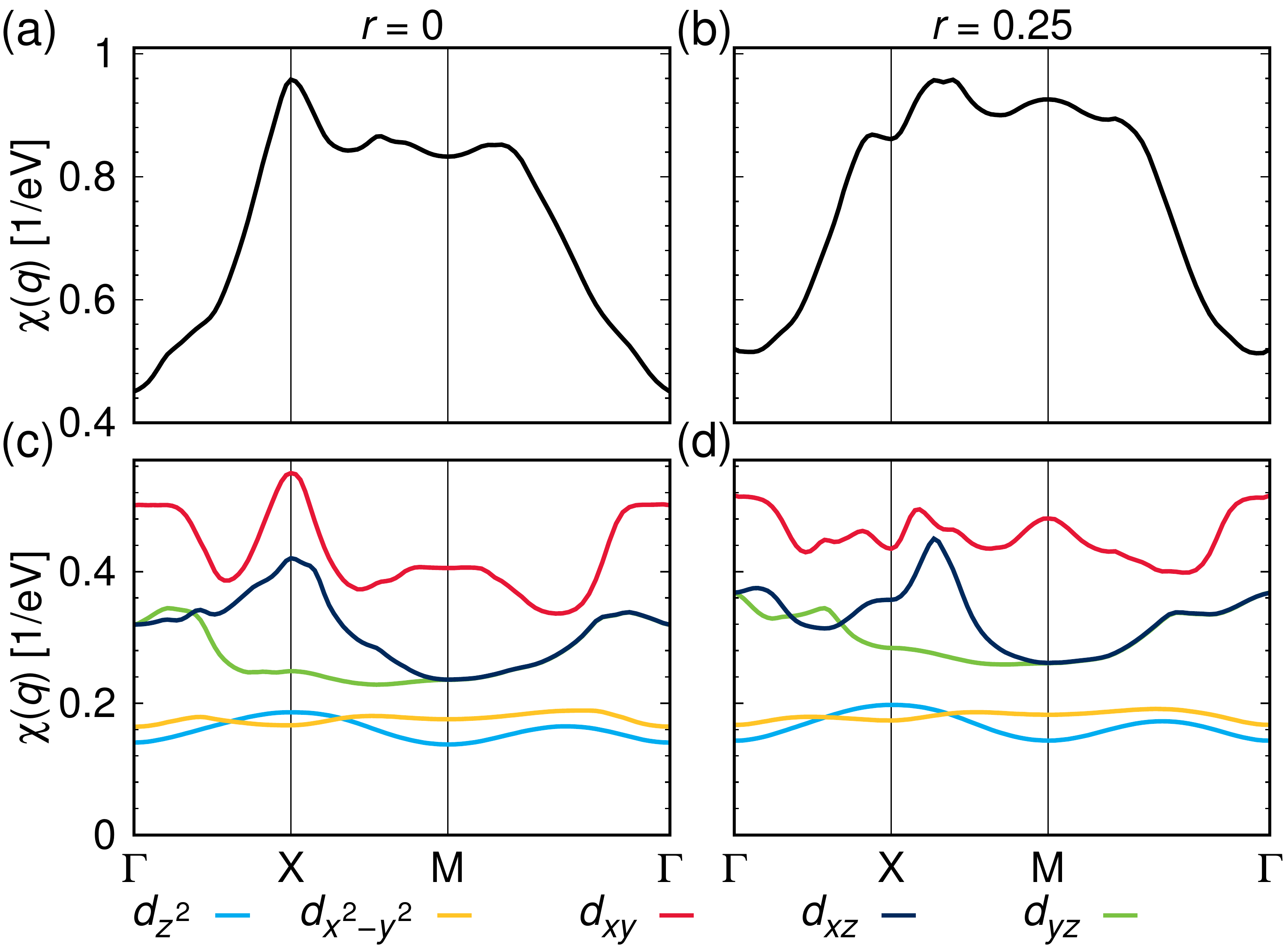}

\caption{(Color online) Summed static susceptibility (upper panel) and its diagonal
  components $\chi_{aa}^{aa}$ (lower panel) in the 8 band tight
  binding model for $r=0.0$ (a,c) and $r = 0.25$ (b,d) in the 1-Fe
  Brillouin zone. The colors identify the Fe 3$d$ states.}
\label{fig:8bandmodelsuscep}
\end{figure}

\begin{figure}[t]
\centering

\includegraphics[width=\linewidth]{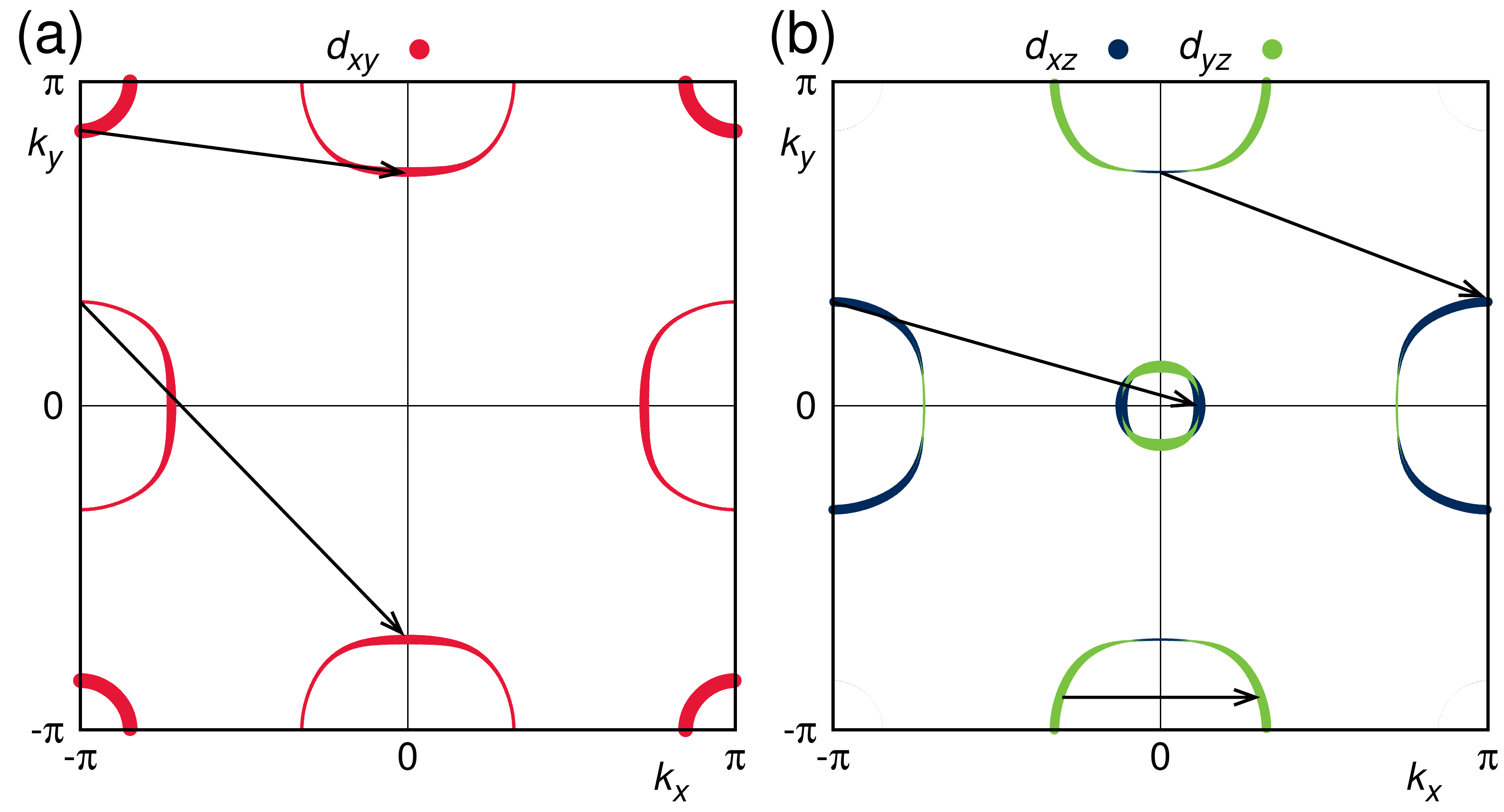}

\caption{(Color online) Fermi surface in the 8 band tight binding model for $r=0.25$ in
the 1-Fe Brillouin zone at $k_z = 0$. Shown are the orbital characters for
the $d_{xy}$ (a) and the $d_{xz/yz}$ (b) orbitals. The arrows represent
dominant interaction vectors identified from peaks in the susceptibility.}
\label{fig:8bandmodelfermisurface}
\end{figure}

\begin{figure*}[tbp]
\centering

\includegraphics[width=\linewidth]{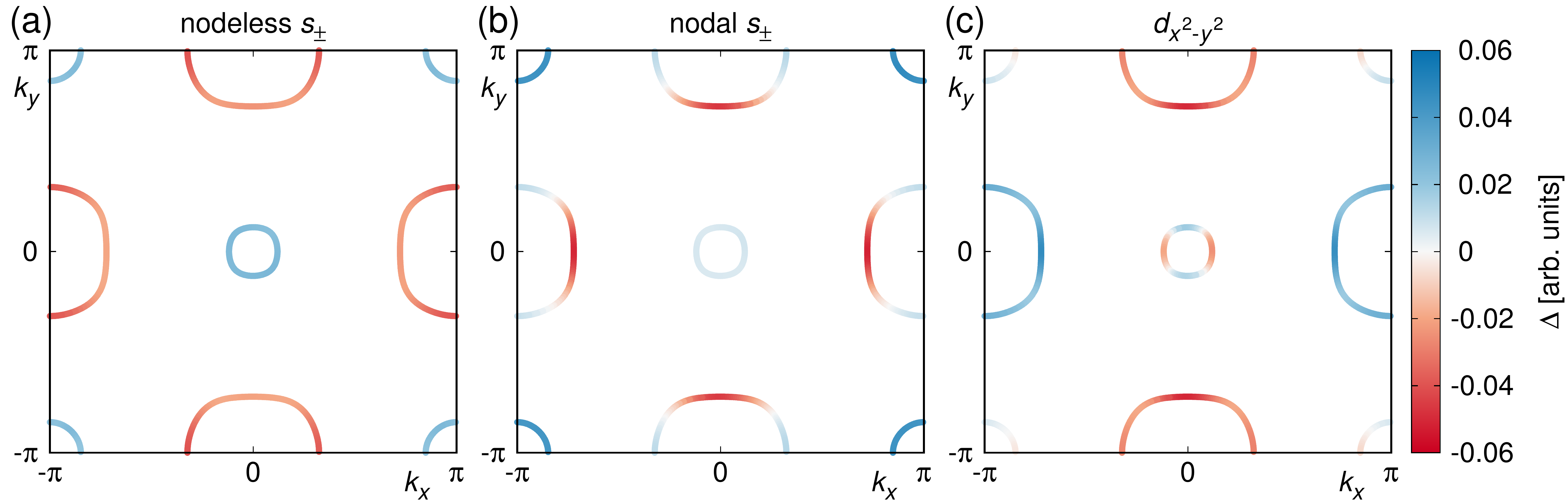}

\caption{(Color online) Solutions of the linearized gap equation on the Fermi surface
  within the 8 band tight binding model for $r=0.25$ in the 1-Fe Brillouin zone at
  $k_z=0$. The relevant instabilities are nodeless $s_\pm$ (a),
  nodal $s_\pm$ (b) and $d_{x^2-y^2}$ (c). We assume spin rotation-invariant interaction
parameters $U=1.35~\mathrm{eV}$, $U'=U/2$, $J=J^\prime=U/4$.}
\label{fig:8bandmodelpairingsolutions}
\end{figure*}

\begin{figure}[tbp]
\centering
\includegraphics[width=\linewidth]{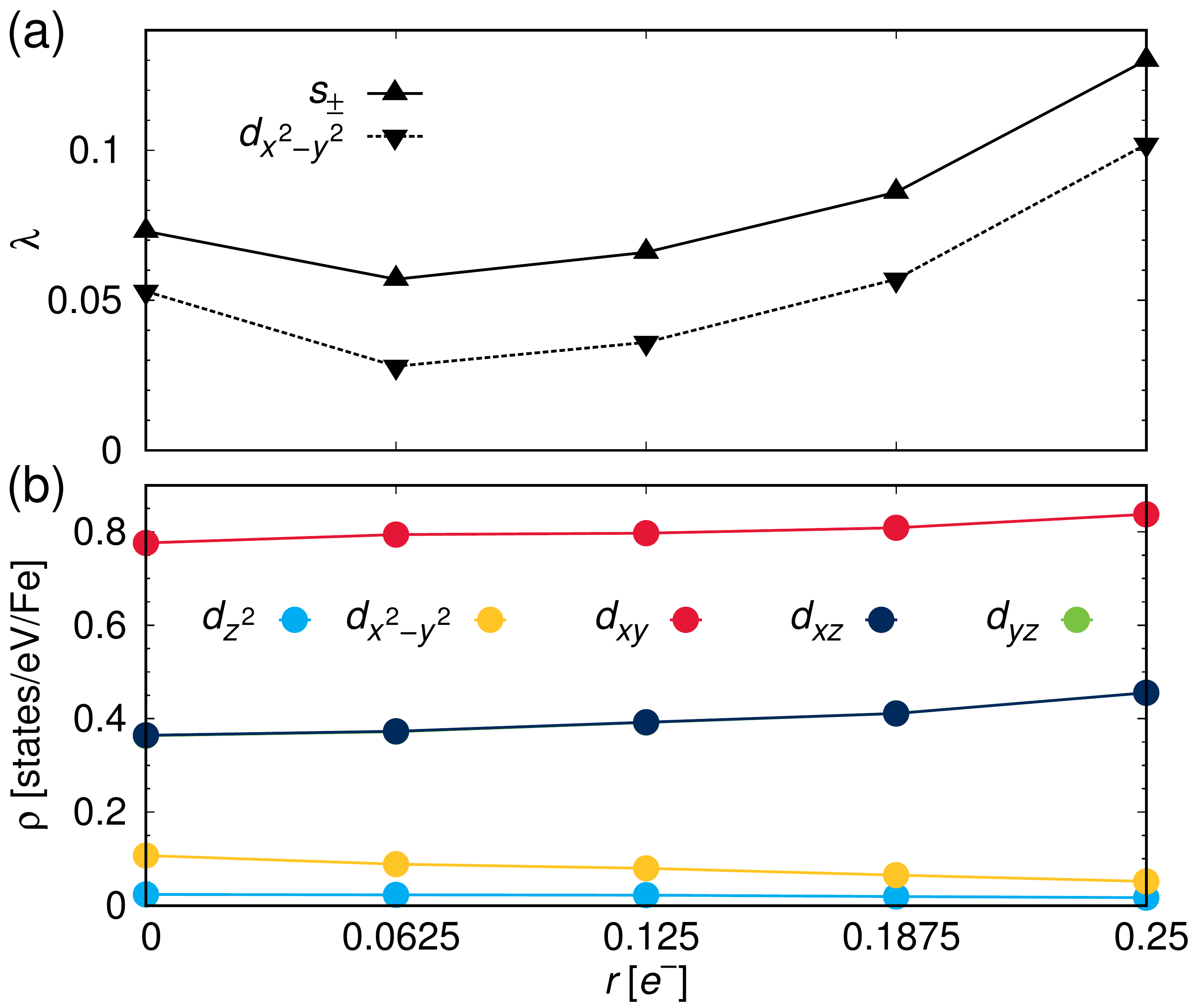}
\caption{(Color online) Trend of the eigenvalues of $s_\pm$ and $d_{x^2 - y^2}$
  solutions (a) and orbital resolved Fe 3$d$ density of states at the
  Fermi level (b) with doping.}
\label{fig:8bandmodelpairingevtrend}
\end{figure}

Next, we investigate the doping dependence of spin fluctuations. The
non-interacting static susceptibility on the high-symmetry path
calculated in the 1-Fe Brillouin zone for $r=0.0$ and $r=0.25$  is shown in
Fig. \ref{fig:8bandmodelsuscep}. The susceptibility
$\chi_{st}^{pq}$ is a four-tensor in orbital indices. The observable
static susceptibility is defined as the sum over all components
$\chi_{aa}^{bb}$. In the undoped compound ($r=0.0$) the structure of the static
susceptibility resembles strongly what is found for materials like
LaFeAsO or BaFe$_2$As$_2$. The electron doping notably shifts the
maximum from $X = (\pi, 0)$ towards $M = (\pi, \pi)$ and the former
valley at $M$ transforms into a peak. The absence of a $(\pi, 0)$ peak
in electron doped compounds suggests why no orthorhombic phase or
stripe-like magnetism have been found in FeSe intercalates 
so far~\cite{IntercalatedSpinFluctuations}.

The shifts of the dominant vectors of spin fluctuations can be
understood from nesting properties and orbital character on the Fermi
surface in the 1-Fe Brillouin zone.  The undoped compound ($r=0.0$) is
dominated by $(\pi, 0)$ nesting of electron and hole pockets, whereas the
electron doped compound ($r=0.25$) (see Fig. \ref{fig:8bandmodelfermisurface})
features scattering between electron and
hole pockets with altered wave vector competing
with scattering between electron pockets. The dominant contributions to
the static susceptibility originate from the $d_{xy}$ and $d_{xz/yz}$
orbitals.

To explore how the superconducting state might depend on interlayer spacing
and doping, we use the 3D version of
random phase approximation (RPA) spin fluctuation
theory~\cite{NJPSpinfluctuationMethod} with Hamiltonian $H=H_{0}+H_{int}$.  Here $H_0$ is the
tight-binding Hamiltonian derived from the DFT calculations using the projective
Wannier function formalism described above, and
$H_{int}$ is the Hubbard-Hund interaction, including the onsite intra (inter)
orbital Coulomb interaction $U$ ($U'$), the Hund's rule coupling $J$ and the
pair hopping energy $J'$. We keep the selenium states in
the entire calculation, but consider interactions only between Fe $3d$
states. 
We assume spin rotation-invariant interaction
parameters $U=1.35~\mathrm{eV}$, $U'=U/2$, $J=J^\prime=U/4$.  The effective
interaction in the singlet pairing channel  is then constructed via the
multiorbital RPA procedure.  Both the original Hamiltonian and effective
interaction are discussed, e.g. in Ref. \onlinecite{HirschfeldReview}.

For all values of electron doping 
(structures $r=0.0$ to $r=0.25$) 
and interlayer spacing  (structures Li$_{0.5}$(NH$_3$)Fe$_2$Se$_2$ 
 and
Li$_{0.5}$(NH$_3$)$_2$Fe$_2$Se$_2$) considered  we find the leading
instability to be of nodeless $s_\pm$ character, while subleading solutions are
of nodal $s_\pm$ and $d_{x^2-y^2}$ type 
(see Fig. \ref{fig:8bandmodelpairingsolutions} for structure $r=0.25$).
These are the leading states expected in the case
of a nearly 2D system with both hole and electron pockets. Repulsive electron-hole
$d_{xz/yz}$ and $d_{xy}$ interactions favor nodeless $s_\pm$ pairing, while interelectron
pocket interactions, orbital weight variations around the Fermi surface, and intraband
Coulomb interactions are known  to frustrate the $s_\pm$ interaction and drive nodal
behavior and eventually $d$-wave interactions when hole pockets
disappear~\cite{Maier_anisotropy,GapEvolutionMaiti}.

We observe that the source of the
moderate quantitative enhancement of $T_c$ with electron doping
lies in an increased
density of states at the Fermi level. For both the $d_{xy}$ and the
$d_{xz/yz}$ orbitals the slope of density of states near the Fermi
level is positive (Fig.~\ref{fig:8bandmodelpairingevtrend} (b))
so that electron doping leads to an enhanced
susceptibility and superconducting pairing strength as the doping
approaches the edge of the hole bands, which appears as a sharp drop
of the $d_{xy}$ DOS~\cite{Supplement}. The small initial decrease of the pairing eigenvalue
at low electron doping (Fig.~\ref{fig:8bandmodelpairingevtrend} (a))
is a consequence of the degraded nesting.

 Alternatively, when we keep the electron doping levels
fixed to the same value and analyze only the interlayer spacing effect
(structures Li$_{0.5}$(NH$_3$)Fe$_2$Se$_2$ with $c=8.1~\mathrm{\AA}$ and 
Li$_{0.5}$(NH$_3$)$_2$Fe$_2$Se$_2$ with $c=10.3~\mathrm{\AA}$),
we find that the Fermi surface turns completely two-dimensional for a $c$-axis length between
$8.1~\mathrm{\AA}$ and $10.3~\mathrm{\AA}$, where $T_c$ saturates in experiment.
Analyzing the susceptibility and superconducting pairing for both structures, we
find no qualitative differences. Quantitatively, the perfectly
two-dimensional Fermi surface of the ammonia rich compound leads to an
increased susceptibility and larger pairing eigenvalue than in the ammonia poor compound.
The increased pairing eigenvalue would correspond to an enhanced $T_c$.

Our calculations show that both increasing electron doping and
lattice spacing contribute to enhancing $T_c$. However, experimentally
 it is found that the ammonia poor
compound (larger electron doping) with smaller $c$-axis
shows a higher $T_c$
(Li$_{0.6}$(NH$_2$)$_{0.2}$(NH$_3$)$_{0.8}$Fe$_2$Se$_2$,
$T_c =
44~\mathrm{K}$) than the ammonia rich compound (smaller electron doping)
with larger $c$-axis 
(Li$_{0.56}$(NH$_2$)$_{0.53}$(NH$_3$)$_{1.19}$Fe$_2$Se$_2$, $T_c =
39~\mathrm{K}$). Therefore, the variations in lattice parameters observed experimentally cannot
be the source of the enhancement of $T_c$. Within our picture,
this leaves only the electron doping level as the controlling parameter.
Hence we conclude that $T_c$ is mainly
controlled by the electron doping level when the Fermi surface is mostly
two-dimensional. Therefore it is unlikely that $T_c$ can be enhanced further by
intercalation of larger molecules.

Summarizing, we investigated  the
Li$_x$(NH$_2$)$_y$(NH$_3$)$_z$Fe$_2$Se$_2$ family of FeSe
intercalates and  found that the FeSe layer is moderately electron
doped. The electron doping moves the Fermi level towards the edge of
the hole-bands, which gives rise to increased superconducting
transition temperatures due to an increase in the density of states at
the Fermi level. We also showed that recently achieved interlayer distances
in FeSe intercalates already produce a two-dimensional Fermi surface, which is optimal
for $T_c$. Further experimental work should therefore concentrate on the charge
doping through the spacer layer.

We would like to thank Stephen J. Blundell and Simon J. Clarke for pointing this 
interesting problem to us and Milan Tomi\'c and Andreas Kreisel for useful 
discussions. We thank the Deutsche Forschungsgemeinschaft for financial support 
through Grant No. SPP 1458. This research was supported in part by the National 
Science Foundation under Grant No. PHY11-25915. PJH and RV thank the Kavli 
Institute for Theoretical Physics at the University of California, Santa Barbara 
for the kind hospitality. PJH acknowledges support by the Department of Energy 
under Grant No. DE-FG02-05ER46236.

\bibliographystyle{apsrev4-1}

\clearpage


\section*{Supplementary material}

# Unified picture of the doping dependence of superconducting transition temperatures in alkali metal/ammonia intercalated FeSe: Supplemental Information

Daniel Guterding,[1, *] Harald O. Jeschke,[1] P. J. Hirschfeld,[2] and Roser Valentí[1]

[1]*Institut für Theoretische Physik, Goethe-Universität Frankfurt,*
*Max-von-Laue-Straße 1, 60438 Frankfurt am Main, Germany*
[2]*Department of Physics, University of Florida, Gainesville, Florida 32611, U.S.A.*

## I. DETAILS OF THE AB-INITIO CALCULATION

For the relaxation of crystal structures we used $6 \times 6 \times 6$ k-point grids within the GPAW code. The structure optimization process was controlled by the Broyden-Fletcher-Goldfarb-Shanno (BFGS) algorithm implemented in the Python atomic simulation environment (ASE)[1]. We checked the remaining forces in the full potential local orbital code FPLO with a $12 \times 12 \times 12$ k-point mesh and found the atomic positions to be sufficiently converged. All further calculations using FPLO were carried out on $12 \times 12 \times 12$ k-point meshes.

We investigated the doping dependence of the electronic structure by means of the virtual crystal approximation (VCA). The VCA can be used to interpolate continuously between the properties of an atom with nuclear charge $Z$ and its neighbors in the periodic table of elements with nuclear charges $Z - 1$ or $Z + 1$. In the study presented here, we fractionally replaced nitrogen ($Z = 7$) by carbon ($Z = 6$). Because of the different valence of nitrogen and carbon this procedure interpolates between charge neutral ammonia ($NH_3$) and a methyl radical ($CH_3^-$). As ammonia ($NH_3$) and methane ($CH_4$) are structurally similar, VCA should provide a sufficient description of the spacer layer.

## II. DETAILS OF THE MODEL CONSTRUCTION

We obtained the low energy Hamiltonian for each doping level from *ab-initio* calculations and subsequent Wannier downfolding using the FPLO Wannier module. We keep all Fe $3d$ and Se $4p$ states in our models. For the 8 Fe supercell this process yields a 64 band tight binding model, which we then treat with the unfolding methods described in the main text. The energy window we used for the fit spans approximately from $-6$ eV to $+2$ eV.

In FPLO a direct unfolding of the DFT band weights is implemented[2]. We used it to check agreement between unfolded DFT band structure and unfolded model band structure in the 2 Fe picture to exclude uncertainties from the Wannierization procedure. The only case where we found small quantitative deviations close to the Fermi level is the $Li_{0.5}(NH_{2.5})Fe_2Se_2$ ($r = 0$) compound, where a nitrogen band at $-1.5$ eV prevents a perfect fit using Fe $3d$ and Se $4p$ states only.

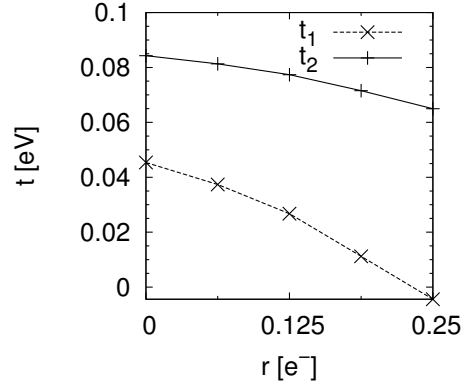

FIG. 1. Nearest and next-nearest neighbor hopping ($t_1$ and $t_2$ respectively) in the Fe $3d_{xy}$ orbital within a 5 band iron only picture.

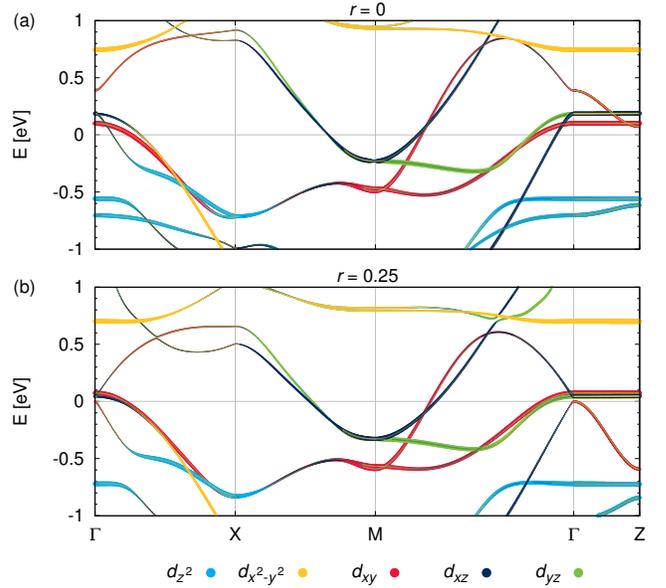

FIG. 2. Band structure in the 16 band tight binding model for $r = 0.0$ (a) and $r = 0.25$ (b) in the 2-Fe Brillouin zone. The colors indicate the weights of Fe $3d$ states.

## III. DETAILS OF THE SUSCEPTIBILITY AND PAIRING CALCULATION

The static susceptibility is calculated on a $30 \times 30 \times 10$ k-point mesh at an inverse temperature of $\beta = 40$ eV$^{-1}$. To calculate the pairing interaction, the susceptibility is

needed at k-vectors that do in general not lie on a grid. Those susceptibility values are obtained using trilinear interpolation of the gridded data.

The gap equation is solved using $\sim 1000$ points on the three-dimensional Fermi surface. Consequently the solution of the gap equation is available only on those points scattered in three-dimensional space. In order to obtain a graphical representation of the gap function on two-dimensional cuts through the Brillouin zone, we used multiscale radial basis function interpolation as implemented in ALGLIB (http://www.alglib.net).

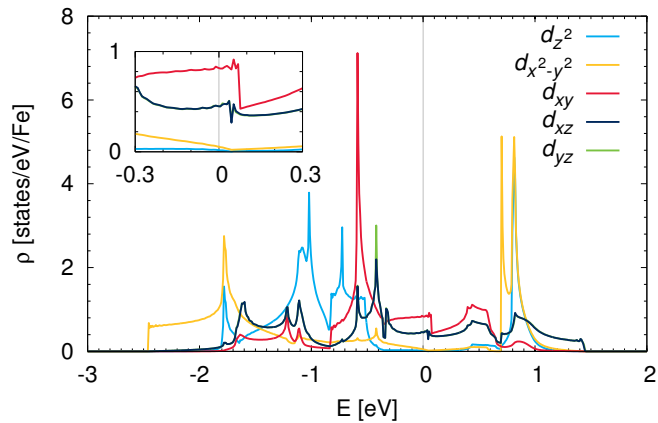

FIG. 3. Orbital resolved Fe $3d$ density of states of the 8 band tight binding model for $r = 0.25$.

## IV. ROLE OF KINETIC FRUSTRATION

An alternative way of understanding the variation in pairing strength with electron doping is provided through the analysis of the tight binding parameters. As mentioned in the main text, we observe a steep decrease of the nearest neighbor hopping with electron doping. To allow easy interpretation of the hopping elements, we constructed 5 band Fe only models by downfolding also the Se $4p$ states. We used an energy window ranging from $-3$ eV to $+2$ eV.

As shown in Fig. 1, the nearest neighbor hopping $t_1$ is affected much stronger by the electron doping than the next-nearest neighbor hopping $t_2$. This behavior has been explained in the literature by the cancellation of direct and indirect hopping paths via the chalcogen atoms[3]. Recently, similar arguments were used to explain the doping dependence of superconductivity in LaFeAsO$_{1-x}$H$_x$ and related materials[4]. In the intercalates we investigated, we find rather small Fe $3d_{xy}$ hoppings because of the large Se height of 1.476 Å, which reduces indirect hoppings via the Se atoms.

The band structure effects of changing individual hopping parameters have been worked out in great detail in Ref. 5. In the materials investigated here, the change in $t_1$ serves to shift the hole bands down (Fig. 2), which gives rise to the increased density of states at the Fermi level.

## V. DENSITY OF STATES

In Fig. 3 we plot the orbital resolved density of states for the 8 band model at $r = 0.25$. The positive slope close to the Fermi level and the edge of the hole bands at $+0.1$ eV are clearly visible.

---



\end{document}